\documentclass[9pt]{article}
\usepackage{spconf,amsmath,graphicx,hyperref}

\usepackage{cite}
\usepackage{algorithm}
\usepackage[numbers,sort&compress]{natbib}
\usepackage{comment}
\usepackage{subfigure}
\usepackage{adjustbox}
\usepackage{caption}
\usepackage{subcaption}

\usepackage{booktabs}
\usepackage{multicol, multirow}
\usepackage{color, colortbl}
\usepackage{xcolor}

\usepackage{makecell}

\title{How Does Instrumental Music Help SingFake Detection?}
\name{\makecell[c]{
Xuanjun Chen$^{1}$, Chia-Yu Hu$^{1}$, I-Ming Lin$^{1}$, Yi-Cheng Lin$^{1}$\thanks{* The first author is with the Graduate Institute of Communication Engineering, National Taiwan University. Email: \texttt{d12942018@ntu.edu.tw}. This work was partially supported by the National Science and Technology Council, Taiwan (Grant no. NSTC 112-2634-F-002-005).}, I-Hsiang Chiu$^{2}$, You Zhang$^{3}$ \\ Sung-Feng Huang$^{4}$, Yi-Hsuan Yang$^{1}$, Haibin Wu$^{5}$, Hung-yi Lee$^{1}$, 
Jyh-Shing Roger Jang$^{1}$
}}
\address{$^1$National Taiwan University
$^2$Carnegie Mellon University
$^3$University of Rochester \\
$^4$NVIDIA
$^5$Independent Researcher
}

\begin{document}
\maketitle
\begin{abstract}
Although many models exist to detect singing voice deepfakes (SingFake), how these models operate, particularly with instrumental accompaniment, is unclear. We investigate how instrumental music affects SingFake detection from two perspectives. To investigate the behavioral effect, we test different backbones, unpaired instrumental tracks, and frequency subbands. To analyze the representational effect, we probe how fine-tuning alters encoders' speech and music capabilities. Our results show that instrumental accompaniment acts mainly as data augmentation rather than providing intrinsic cues (e.g., rhythm or harmony). Furthermore, fine-tuning increases reliance on shallow speaker features while reducing sensitivity to content, paralinguistic, and semantic information. These insights clarify how models exploit vocal versus instrumental cues and can inform the design of more interpretable and robust SingFake detection systems.
\end{abstract}
\section{Introduction}

Advancements such as VISinger \cite{zhang2022visinger} and DiffSinger \cite{liu2021diffsinger}, have advanced singing voice generation. 
However, these developments raise concerns for artists, record companies, and publishing houses. 
The potential for unauthorized synthetic imitations of a singer's voice threatens the commercial value of the artists. 
Recently, singing-voice deepfake (SingFake) detection datasets \cite{zang2023singfake, zang24_interspeech} and challenges \cite{zhang2024svdd_slt} have heightened awareness of the need to safeguard artists’ rights and revenues while ensuring the integrity of musical productions. 

Before digging into SingFake detection, it is natural to ask how it differs from deepfake speech detection. 
Two factors may explain the domain gap between these tasks.  
First, speech is spontaneous and unconstrained \cite{li24na_interspeech}, whereas singing must adhere to melodic and rhythmic structures \cite{wang2023asru} that intentionally modify the pitch and duration of phonemes. 
Second, singing encompasses a broader timbral range and a wider range of artistic expression. 
These musical properties can mask or transform the subtle artifacts that speech-deepfake countermeasures (CM) typically exploit, rendering such methods less effective for singing-voice deepfake scenarios. 
Additionally, empirical studies have demonstrated that models trained on ASVspoof 2019 \cite{wang2020asvspoof} exhibit poor performance in SingFake \cite{zang2023singfake} detection. 
Recent work further demonstrates that, beyond direct training on singing-voice deepfake data \cite{anmol2024slt_singfake, zhang2024_singfake_xwsb, liu2025nes2net}, integrating instrumental music \cite{chen24o_interspeech} can further enhance SingFake detection performance. 

In deepfake speech research, researchers have increasingly turned their focus to Explainable AI (XAI) to understand and refine deepfake speech countermeasure (CM) systems. These methods have identified various spoofing artifacts by analyzing silence regions \cite{muller2021speech}, background noise \cite{salvi2024listening}, frequency bands, and speech formats \cite{Tak0NET20-odyssey}. Moreover, gradient-based saliency maps to show that buzziness and rhythmic patterns are critical CM cues \cite{halpern20_odyssey}, while SHAP analyses have uncovered generator-specific artifacts \cite{ge2022explaining, ge22_odyssey}. Visualization techniques like Grad-CAM have helped highlight key discriminative areas \cite{zhang2023impact}, and more recently, investigations into partial-fake audio have revealed how a CM's focus shifts between correct and incorrect predictions \cite{liu24m_interspeech}. 
When considering SingFake, singing resembles speech yet remains uniquely tied to instrumental accompaniment, with vocals naturally following rhythm and melody. 
Although work in the related domain of AI-generated music has used Fourier analysis to find artifacts \cite{afchar2025fourier}, no work has yet explored the relationship between instrumental music and SingFake.

Although instrumental music has been shown to enhance SingFake detection \cite{chen24o_interspeech}, the mechanisms underlying this improvement remain unclear. 
Therefore, this paper aims to investigate the overarching question: \textit{``How Does Instrumental Music Help SingFake Detection?''} 
To answer this question, our work provides three main contributions:

\begin{itemize}

\item We conduct a series of comprehensive behavioral analyses to demonstrate that, in the context of SingFake detection, instrumental music functions primarily as data augmentation rather than as a source of intrinsic musical cues such as melody or rhythm. 
\item Through subband analysis, we identify that current models are heavily reliant on low-frequency information within the singing voice itself, while being largely insensitive to cues from the instrumental track.
\item We offer novel representational insights by probing the SingFake model's encoders before and after fine-tuning. 
We find that this fine-tuning process enhances sensitivity to speaker-specific features while diminishing capabilities related to deeper content, paralinguistic, and semantic understanding. 
\end{itemize}

\section{Experimental Setup}
To systematically investigate the role of instrumental music in SingFake detection, we designed and evaluated experiments using four representative CM models. 
This set was chosen to ensure our conclusions are not dependent on a single architecture. The models include: Spec-ResNet \cite{resnet18}, which processes spectrograms with a ResNet18 architecture; AASIST \cite{jung2022aasist}, which operates on raw waveforms using graph neural networks; W2V2-AASIST \cite{tak2022automatic}, which combines AASIST with the Wav2Vec2 \cite{baevski2020wav2vec} front-end; and SingGraph \cite{chen24o_interspeech}, which fuses Wav2Vec2 and the MERT musical model \cite{li2024mert}. While the first three were originally developed for deepfake speech, SingGraph was designed specifically for SingFake detection. It is noteworthy that Wav2Vec2 and MERT are pre-trained on large-scale speech and music datasets, respectively.

All CMs take audio as input and output a binary prediction indicating whether a singing voice is bona-fide or a deepfake. Depending on the configuration, the inputs were either vocals only or vocal–instrumental pairs separated using Demucs \footnote{\url{https://github.com/facebookresearch/demucs}}. We applied RawBoost \cite{tak2022rawboost} augmentation to each waveform input during training. All experiments were conducted on the SingFake dataset \cite{zang2023singfake}, which comprises 28.93 hours of real and 29.40 hours of fake song clips split into training, validation, and the T01 test set. We evaluated performance only on T01, as T02–T04 are out-of-domain subsets that could introduce confounding factors. 
The training set contained data from 12 singers, while the validation set included 4 unseen singers. Performance was measured by Equal Error Rate (EER).

\section{Behavioral Effect Analysis}
We analyze SingFake detection from two aspects: behavioral, to reveal what cues models rely on, and representational, to uncover how fine-tuning reshapes encoder feature spaces. 
We begin by examining the behavioral effect of instrumental music on SingFake detection. Specifically, we investigate whether it benefits different backbones consistently, whether unpaired instrumental music can aid detection, and which frequency subbands contribute the most to performance.

\subsection{Do Different Backbones Consistently Benefit from Instrumental Music?}
Our results show that \textbf{instrumental music consistently improves detection performance across different model architectures}, confirming its general benefit for the SingFake task.
To examine this, we compare four backbones: Spec-ResNet, AASIST, W2V2-AASIST, and SingGraph, under two input conditions: vocal only versus vocal with instrumental music. 
This setup allows us to examine whether instrumental music helps improve detection performance. 
In Table~\ref{tab:different_model}, across three models, adding instrumental music consistently lowers the EER compared to using \texttt{Vocal-Only} (e.g., Spec-ResNet from 11.78 to 10.82, AASIST from 5.24 to 5.51, SingGraph drops from 5.24 to 3.77), showing that accompaniments indeed improve detection performance.

\begin{table}[t]
\renewcommand{\tabcolsep}{5.5pt}
\fontsize{9}{11}\selectfont
\caption{Comparison across different models (EER\%).}
\centerline{
\begin{tabular}{l c c}
\toprule
 Model & \texttt{Vocal-Only}    & \texttt{Vocal-Inst.}   \\
\midrule

Spec-ResNet \cite{resnet18} &  11.78\%  & 10.82\% \\
AASIST \cite{jung2022aasist} &    6.78\% & 6.28\% \\
W2V2-AASIST \cite{tak2022automatic} &   5.24\%  &  5.51\% \\   
SingGraph \cite{chen24o_interspeech} &    5.24\% &  3.77\%  \\

\bottomrule
\end{tabular}
}
\label{tab:different_model}
\end{table}

\begin{table}[t]
\renewcommand{\tabcolsep}{3pt}
\renewcommand\arraystretch{1.3}
\fontsize{9}{11}\selectfont
\caption{Comparison of different input modalities (EER\%).}
\centerline{
\begin{tabular}{l c c c c c}
\toprule
Input Modality &  Baseline &  Speech  &  Noise  &  Music  &  Reverb  \\

\midrule
\texttt{Vocal-Inst.}  &  3.77\%  & 3.97\% & 4.08\%  & 3.77\%  &  3.97\%  \\
\texttt{Vocal-Noise}  &  3.77\%  & 3.47\% & 3.77\%  & 3.31\%  &  3.77\%  \\
\texttt{Vocal-Music}  &  3.77\%  & 3.77\% & 3.81\%  & 3.51\%  &  3.77\%  \\
\bottomrule
\end{tabular}
}
\label{tab:unpair_comparison}
\end{table}

\begin{figure*}[t]
\centering
\includegraphics[width=17.5cm]{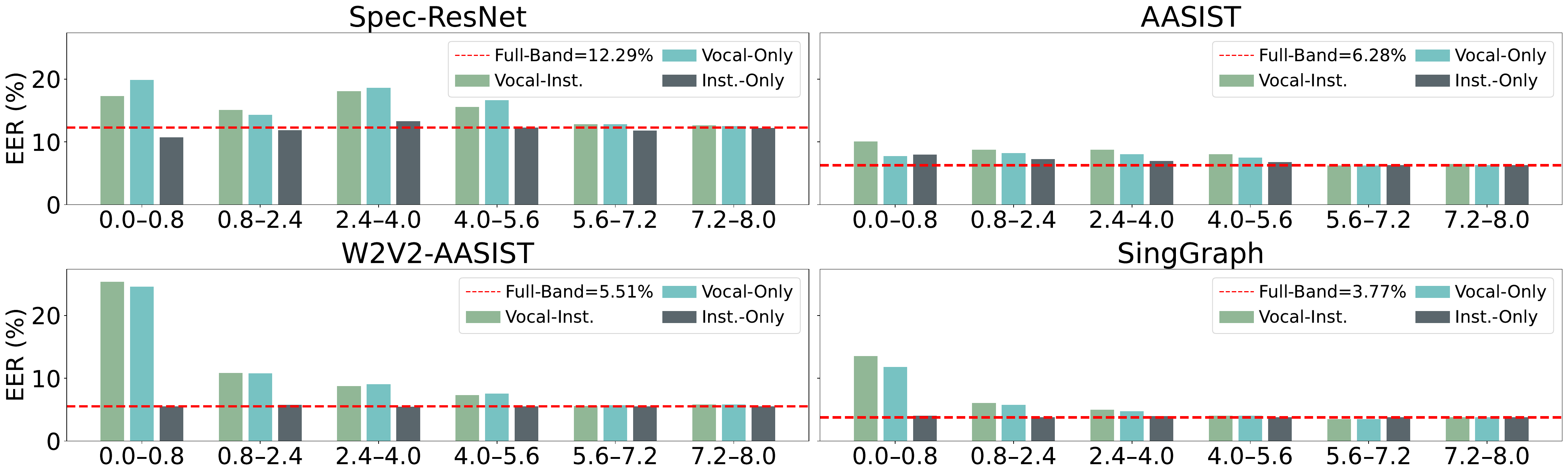}
\caption{Performance under band-stop filtering for three inputs: \textit{Vocal-Inst.}, \textit{Vocal-Only}, and \textit{Inst.-Only}, while ``Full-Band'' denotes all-pass filtering. The x-axis indicates the frequency ranges (kHz) to be filtered out.}
\label{fig:subband-analysis}
\end{figure*}

\subsection{Can Unpaired Instrumental Music Aid Detection?}

Contrary to the intuition that models might require original musical accompaniment cues, \textbf{we found that alignment between the vocal and instrumental tracks is not necessary for performance improvement}. 
In fact, using unpaired music or even noise can be as effective or even better. 
This suggests the accompaniment serves as data augmentation rather than a source of complementary information.
Therefore, we compared performance using mismatched (unpaired) instrumental tracks or noise against the original paired accompaniments.
Since SingGraph achieved the best performance in Table \ref{tab:different_model} and appeared to leverage instrumental music most effectively, we used this model for our analysis.
We assessed the influence of accompaniment by replacing the original tracks with segments from the MUSAN dataset \cite{musan2015} containing music, speech, and noise, to control background conditions.

Table \ref{tab:unpair_comparison} reports SingGraph’s results under different training–testing modality pairs. The rows correspond to training setups: \texttt{Vocal-Inst.} (original pairs), \texttt{Vocal-Noise} (paired with noise), and \texttt{Vocal-Music} (paired with unrelated music). The columns denote test conditions: Baseline, Speech, Noise, Music, and Reverberation. This layout highlights how training modality and test interference jointly affect performance. Under the baseline, all three setups perform similarly. Surprisingly, \texttt{Vocal-Noise} and \texttt{Vocal-Music} match \texttt{Vocal-Inst} despite lacking alignment. With test-time augmentations, \texttt{Vocal-Inst.} consistently underperforms, and \texttt{Vocal-Noise} slightly outperforms \texttt{Vocal-Music}. 
These results contradict the intuitive expectation that aligned accompaniment helps, and instead suggest that mismatched backgrounds act as data augmentation, improving robustness to interference. 

These results directly answer whether the model exploits complementary instrumental music cues. 
Replacing the true accompaniment with unrelated music or noise maintains or slightly improves performance, indicating that the model does not rely on any deepfake-specific information from the original instrumentals. 
Instead, these "unpaired" tracks simply act as a stronger augmentation. By varying the background conditions, the model is compelled to ignore all accompaniment content and concentrate solely on invariant vocal features. In other words, rather than harnessing complementary cues from the music, the network uses the instrumental track purely to regularize its focus on discriminative vocal signals.

\subsection{Which Frequency Subbands Contribute Most?}

Vocals and musical harmonics occupy different subbands, potentially contributing differently to detection. 
Our subband analysis revealed that models \textbf{rely almost entirely on low-frequency cues (0–0.8 kHz) from the singing voice}, while being largely insensitive to the instrumental track. 
To investigate this, we applied 4th-order Butterworth band-stop filters to the six frequency ranges of the paired vocal and instrumental music tracks, then fed the filtered audio into each model as the new input. The models were trained with \texttt{Vocal-Inst.} setup and not further retrained; filtering was applied only at inference. 
Each system was then tested with \textit{Vocal-Only} (only vocal track filtered), \textit{Inst.-Only} (only instrumental track filtered), and \textit{Vocal-Inst.} (both tracks filtered) inputs to measure the resulting performance degradation and identify the most critical frequency regions for detection (Figure~\ref{fig:subband-analysis}).

We observe two key trends in Fig. \ref{fig:subband-analysis}.
First, masking the 0–0.8 kHz band sharply increases error rates for all models, most dramatically for SingGraph and W2V2-AASIST, revealing a strong dependence on low-frequency content. 
Second, when we mask low frequencies in the vocal track (for both \textit{Vocal-Only} and \textit{Vocal-Inst.} inputs), performance degrades substantially. 
By contrast, masking bands in \textit{Inst.-Only} inputs have little effect. 
Together, the results show models rely on vocal's low-frequency cues rather than instrumental ones. 
These trends can be explained by two factors. First, both SingGraph and W2V2-AASIST rely on Wav2Vec2 encoders pretrained on speech, which emphasizes the 150–2000 Hz range \cite{wang22_odyssey}. Therefore, removing the lowest band (0–0.8 kHz) eliminated crucial information that these models expect, making their representations less informative. 
Second, masking the same bands in accompaniment had little effect, as instrumental tracks contain no deepfake-specific artifacts for the models to exploit. 
Once the primary vocal cues were absent, instrumental changes failed to influence the performance. 

These behavioral findings lead to a deeper question: why do the models learn to rely on these specific vocal cues, and how does fine-tuning for SingFake detection alter their internal representations? To answer this, we next turn to a representational analysis of the encoders. 

\begin{figure*}[t]
\centering
\includegraphics[width=17.5cm]{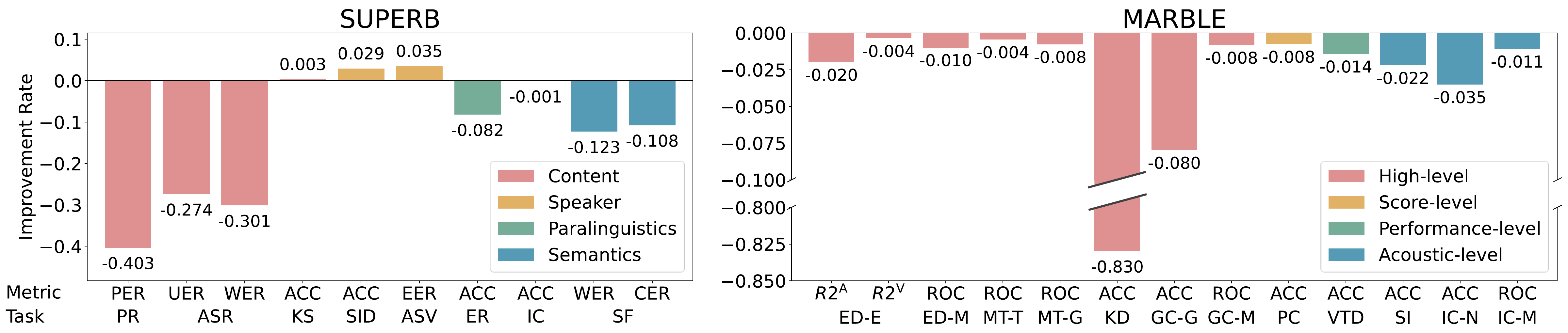}
\caption{Probing analysis of the self-supervised learned encoders for the vocal and instrumental music features. In the MARBLE, each task is denoted by an abbreviation, where the letter after the hyphen is derived from the corresponding dataset name.}
\label{fig:encoder-analysis}
\end{figure*}

\section{Representational Effect Analysis}
Having established the models' behavioral patterns, we probe their internal mechanisms to understand the representational shifts that occur during training. 
Our analysis reveals a trade-off: \textbf{fine-tuning on SingFake detection makes the encoders highly sensitive to shallow, speaker-specific features, while simultaneously degrading their ability to process deeper content, paralinguistic, and musical information.}

To arrive at this conclusion, we benchmarked the vocal (Wav2Vec2) and instrumental (MERT) encoders before and after fine-tuning using two established suites: SUPERB \cite{yang2021superb} for speech representations and MARBLE \cite{yuan2024marble} for music. SUPERB assesses representations across tasks related to content, speaker identity, and paralinguistic features, while MARBLE evaluates music features from acoustic to semantic levels. Due to space limitations, the specific details of the 8 SUPERB and 12 MARBLE tasks we used are not described here; for other details, please refer to SUPERB \footnote{\href{https://superbbenchmark.github.io}{https://superbbenchmark.github.io}} and MARBLE \footnote{\href{https://github.com/a43992899/MARBLE}{https://github.com/a43992899/MARBLE}} benchmark paper. 
Our procedure involved extracting baseline embeddings from the pretrained encoders, then fine-tuning them within the SingGraph model on the SingFake task. We then extracted new embeddings and re-evaluated them on the benchmarks. This pre- vs. post-fine-tuning comparison revealed which representations were retained or diminished. To quantify this, we defined an improvement rate as the relative difference between the post-fine-tuning score $S$ and the original baseline $O$. For accuracy metrics, the rate was $(S - O)/O$; for error metrics, it was $(O - S)/O$, ensuring a positive value always indicated a gain.

As shown in Figure \ref{fig:encoder-analysis}, fine-tuning on SingFake generally diminished the encoders' original representational capacity. On the SUPERB benchmark, content-related tasks like Phoneme Recognition (PR) and Automatic Speech Recognition (ASR) dropped sharply (up to 40\%), while prosodic and semantic tasks saw moderate declines. This suggested that paralinguistic features were sacrificed for detection cues. In contrast, speaker-related tasks like Speaker Identification (SID) and Automatic Speaker Verification (ASV) improved, indicating a shift toward speaker-specific sensitivity. On the MARBLE benchmark, performance on most music tasks declined slightly, but Key Detection (KD) degraded dramatically, implying a loss of global harmonic context. Overall, the SingFake objective favored shallow artifact cues at the expense of deeper linguistic and musical representations. 

The degradation of the music encoder's capacity was unsurprising, as the instrumental tracks served only as data augmentation. The vocal encoder, however, exhibited a contrasting pattern. We hypothesize that the discrepancy between improved speaker-related performance and degraded content-related results stems from residual accent cues in the synthetic speech. Each singer’s voice bears subtle markers of their native language. Even when a voice conversion system generates singing in a non-native tongue (e.g., a native Chinese speaker performing in English), underlying pronunciation patterns, vowel quality, and intonation contours often persist. Although the output may sound high-fidelity, these lingering accent artifacts can betray its synthetic origin and provide reliable signals for the detection model. 

A broader impact of our work is that it moves beyond prior analyses of deepfake speech, such as the CCA-based study of style–linguistic mismatches in \cite{NEURIPS2024_7d6930fd}, which focused only on coarse speaker–content differentiation but largely overlooked semantic and paralinguistic information. 
In contrast, we systematically quantify how fine-tuning on SingFake shifts encoder representations across speech properties and music properties, by comparing on self-supervised tasks from SUPERB and MARBLE benchmarks. Moreover, by experimentally isolating individual audio properties, we reveal which attributes most strongly drive deepfake detection. 

\section{Conclusion}
We draw three main findings in response to the question ``How does instrumental music help SingFake detection?''
First, instrumental music functions primarily as a form of data augmentation rather than supplying meaningful musical cues. 
Second, behavioral analyses show that models depend almost entirely on low-frequency vocal information, while third, representational probing reveals that fine-tuning amplifies speaker-related features but diminishes sensitivity to content, paralinguistic, and semantic cues. 
Taken together, these findings indicate that instrumental music sharpens the model’s reliance on vocals, offering guidance for designing more interpretable and robust SingFake detectors.

\bibliographystyle{IEEEbib}
\bibliography{strings,refs}

\clearpage

\end{document}